\documentclass[journal = aamick, manuscript = article, email = false]{achemso}
\setkeys{acs}{keywords = true, email = true}

\usepackage{times, multirow}
\usepackage{caption, float, geometry, natbib, setspace, xkeyval}
\usepackage[table]{xcolor}

\usepackage{graphicx, svg, xr, pdfpages, wrapfig, soul}
\usepackage[switch, columnwise]{lineno}
\usepackage{amsmath, cancel}
\usepackage{chemformula} 
\usepackage[T1]{fontenc} 
\usepackage{ebgaramond-maths}
\usepackage{achemso, booktabs}
\usepackage{setspace}

\title{Spectrum Selective Interfaces and Materials towards Non-photothermal Saltwater Evaporation: Demonstration with a White Ceramic Wick}
\author{Navindra Singh}
\affiliation{Material Science and Engineering Program. University of California at Riverside, 900 University Ave, Riverside, California, 92521, United States}
\author{James Leung}
\affiliation{Mechanical Engineering Department. University of California at Riverside, 900 University Ave, Riverside, California, 92521, United States}
\author{Ji Feng}
\affiliation{Mechanical Engineering Department. University of California at Riverside, 900 University Ave, Riverside, California, 92521, United States}
\author{Alma K. González-Alcalde}
\affiliation{Mechanical Engineering Department. University of California at Riverside, 900 University Ave, Riverside, California, 92521, United States}
\author{Arial Tolentino}
\affiliation{Mechanical Engineering Department. University of California at Riverside, 900 University Ave, Riverside, California, 92521, United States}
\author{David Tuft}
\affiliation{Mechanical Engineering Department. University of California at Riverside, 900 University Ave, Riverside, California, 92521, United States}
\author{Juchen Guo}
\affiliation{Material Science and Engineering Program. University of California at Riverside, 900 University Ave, Riverside, California, 92521, United States}
\author{Luat T. Vuong}
\affiliation{Mechanical Engineering Department. University of California at Riverside, 900 University Ave, Riverside, California, 92521, United States}
\alsoaffiliation{Material Science and Engineering Program. University of California at Riverside, 900 University Ave, Riverside, California, 92521, United States}
\email{luatv@ucr.edu}

\begin{document}
\maketitle
\begin{abstract}
    Most solar desalination efforts are photothermal: they evaporate water with ``black'' materials that absorb as much sunlight as possible. Such ``brine-boiling'' methods are limited by the high thermal mass of water, i.e., its capacity to store and release heat. Here, we study the light-enhanced evaporation by a hard, white, aluminum nitride wick, and propose a route to selectively target salt-water bonds instead of bulk heating via deep-UV interactions. Through experiments and analyses that isolate the effects of light absorption and heating in aluminum nitride, we provide experimental evidence of a light-driven, spectrum-selective path to non-photothermal saltwater evaporation. Leverage of these light-matter interactions in white ceramic wicks may achieve low-cost, low-energy desalination, reduce the heat island effects of traditional solar technologies, and contribute to future cooling technologies where drought is also a concern.
\end{abstract}
\section{\normalsize Introduction}
\par Water is an essential resource that increasingly requires treatment. Oceans, endorheic lakes, geothermal brines, and industrial effluent contain dissolved chemicals such as minerals, salts, precious critical elements, and heavy metals \cite{Ogunbiyi2021, Bello2021, McKibben2025}. Many water purification and separation technologies aim to achieve zero-liquid discharge (ZLD) \cite{Finnerty2017, Abimbola2021, Tong2016}, i.e., the full separation of the solute and water so that both may be reused without detrimental environmental impacts. Towards ZLD, there is significant interest in robust and economical strategies that treat high volumes of industrial brine with low capital and operation costs \cite{Tong2016}. 

\par To offset the operation energy costs, solar technologies are widely considered. However, most solar approaches to water treatment continue to be resource intensive (even as they re-

\noindent duce carbon emissions and circumvent the need for grid connectivity) \cite{Menon2023}. Direct solar evaporation generally involves ``black'' materials and material structures and interfaces that absorb broadband light, convert it to heat, and in cases for the highest-efficiency systems, recycle the thermal radiation \cite{Mascaretti2023, Zhu2019, Zhang2020, Xu2020, Yang2024, Zhang2021a, Schmid2022}. Unfortunately, photothermal water separation is an energy-intensive evaporative process due to the specific heat and latent heat of vaporization of water \cite{Generous2020, Sharqawy2010, Tong2016}. Additionally, local heat trapping and heat island effects are exacerbated by the waste heat produced by photothermal water treatment \cite{Wagner2022, BarronGafford2016}.


\par
\begin{figure}[h]
    \centering
    \includegraphics[width=0.45\textwidth]{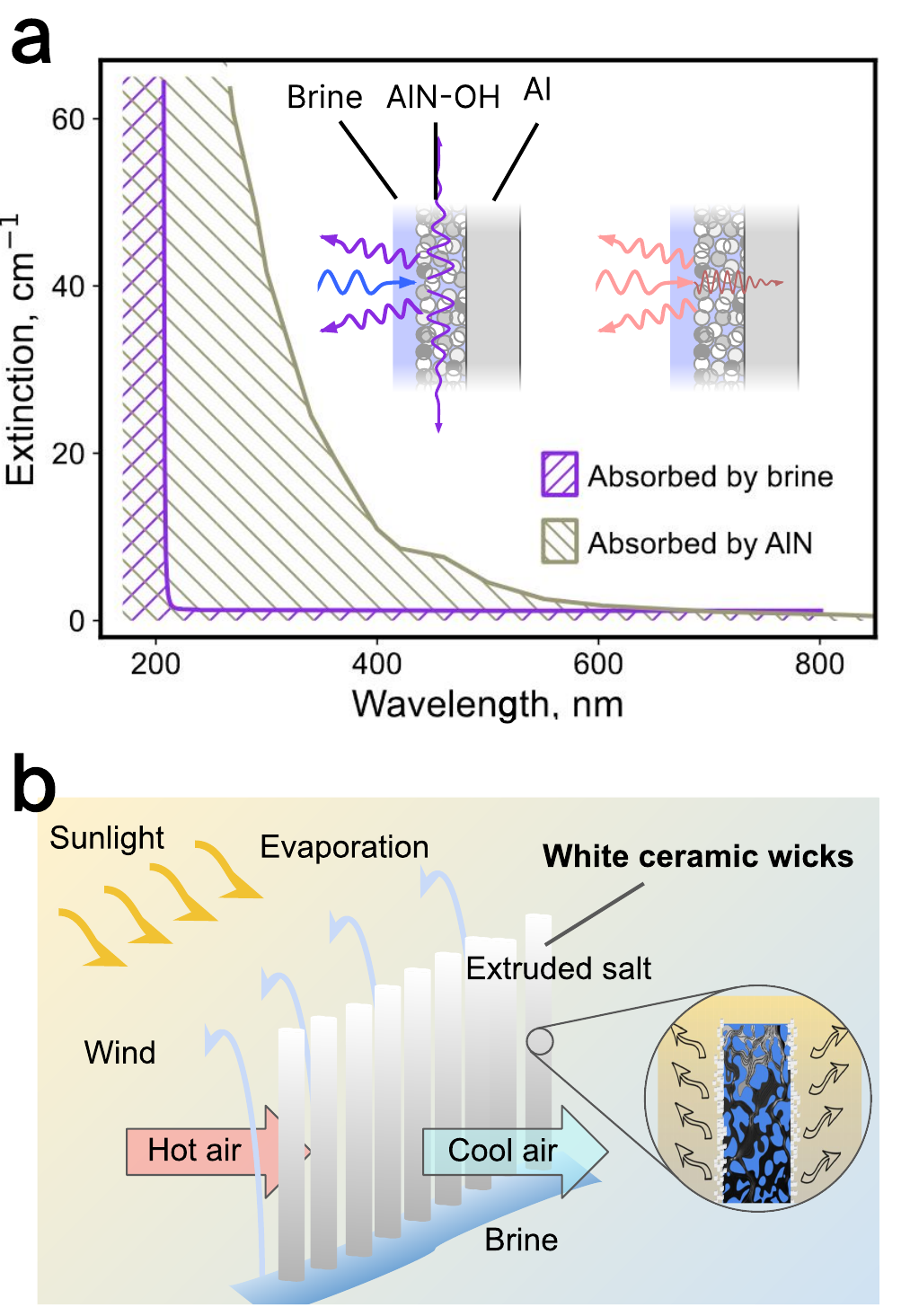}
    \setcitestyle{numbers}
    \caption{(a) Extinction spectra of saltwater brine and AlN. (AlN spectra is  adapted with permission from Ref. [\cite{Nagai2010}] . Copyright Optical Society of America, 2010). Inset: Schematic showing  blue light absorption and upconversion in AlN-OH (left) and red light transmission and absorption in Al (right).  (b)  Illustration of a  ceramic wick in an outdoor saltwater swamp cooler.}
    \label{motivation}
\end{figure}

 Here, we demonstrate a ZLD saltwater separation approach via spectrum-selective light-enhanced evaporation from a white, porous ceramic wick [Fig. \ref{motivation}(a) ]. The ceramic wick serves as a capillary-driven interface and enables high surface-area interactions with the brine. In this demonstration, AlN is well-suited to this spectral approach due to its peak absorption in the deep UV [Fig. \ref{motivation} (a)]. Deep UV light is absorbed by  salt water and not  fresh water, which implies that deep-UV light will selectively target and enhance the evaporation of  salt water without adding excess heat or energy to bulk water. Meanwhile, water is transparent to visible and near-UV light, regardless of its salt content. 


\par Evaporation rates dramatically increase with short-wavelength illumination, which is an indication of targeted salt-water interactions with aluminum nitride. Violet-light illumination achieves 4-10$\times$ higher evaporation enhancement compared to orange and IR light even though at each wavelength, a similar fraction of light is absorbed. With violet light, or with photon energy approximately half that of salt-water bonds, we achieve a light-induced enhanced brine evaporation efficiency of 90-150\%. However, thermal leakage or loss from the wick is responsible for one-third of the evaporation: the high net efficiencies indicate a light-induced saltwater evaporation rate above the photothermal limit. We propose a pathway for non-photothermal evaporation that involves the absorption of visible light and emission of deep UV light by AlN. Such upconversion processes may be outsized with AlN \cite{Hoffmann2013, Aguilar2019, Troha2016} and facilitated by surface water \cite{Neuweiler2019, Tu2023, Lv2024}. The controlled presence of photon upconversion would enable low energy, high efficiency solar desalination, which we discuss further in Sec. \ref{disc}.

\par These results identify opportunities to operationalize hard, white ceramic wicks in light-powered ZLD systems not only for water separation and mineral harvesting \cite{Chen2023}, but also for evaporative cooling. AlN is inexpensive (readily available in powder form) and biocompatible and nontoxic (as a group III nitride, AlN does not promote cell adhesion to its surface) \cite{Du2023, Kobayashi2006, Chasserio2008, Berg2017}. AlN is one among many white hydrophilic materials with deep-UV absorption and potential for deep-UV upconversion (non-centrosymmetric crystal structure) that warrant further study for light-actuated salt water separation opportunities. We believe that knowledge of the spectral dependence of the interfacial energy-transfer may be central to minimizing heat island effects or maximizing convection or fluid dynamics \cite{Cho2016, Abdelsalam2024, Zhang2022}. The light-powered ZLD system mirrors many capillary-based, bio-inspired approaches to desalination \cite{Wang2020c, Abdelsalam2024, Zhu2025, Mi2019}. Moreover, ceramic wicks may be reused or serve as a structural material in hot, sunny regions prone to drought without contributing to further water shortage. 
In Fig. \ref{motivation} (b), we illustrate such a contemporary vertical outdoor cooling shelter \cite{Davis2015}-- an opportunity to efficiently cool with the combined effects of both light and wind. 

\begin{figure}[t]
    \centering
    \includegraphics[width=\textwidth]{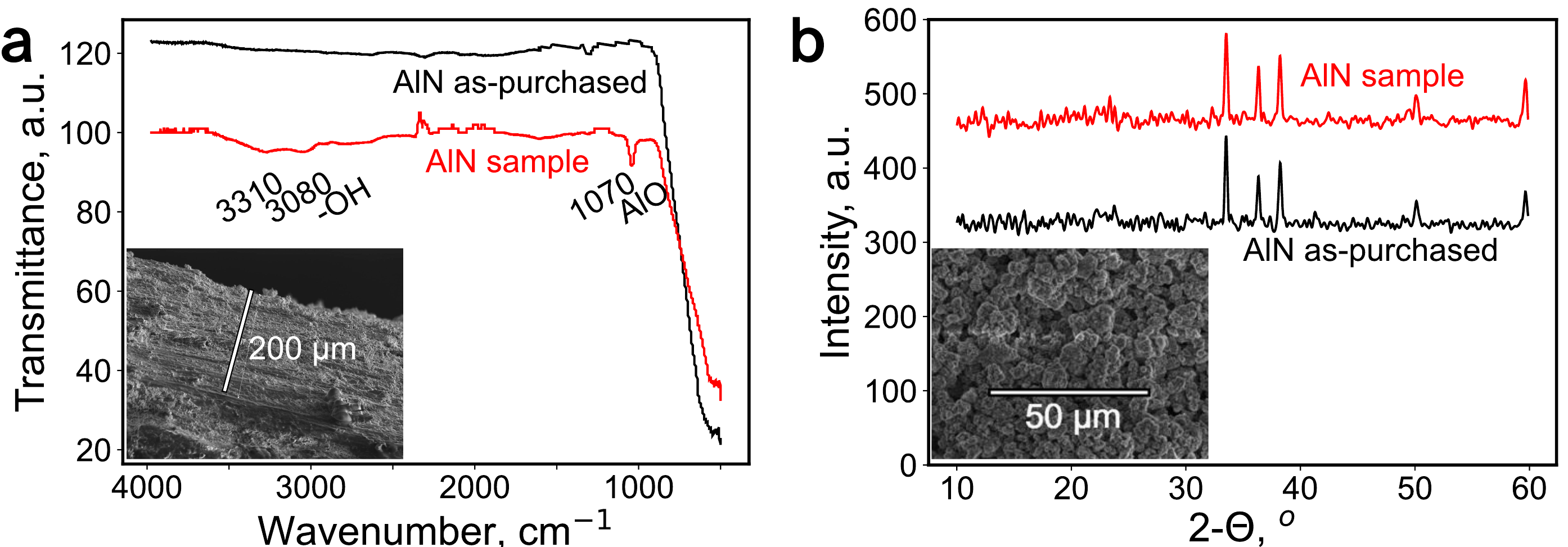}
    \caption{AlN wick characterization. (a) FTIR and (b) XRD spectra of the sample pre- and post-fabrication, with inset SEM photos showing porosity and thickness of AlN slab. The fabricated AlN sample (red) shows similar chemical and structural composition to the as-purchased AlN (black) with the exception of significant -OH as seen in the FTIR.}
    \label{aln_props}
\end{figure}

\section{\normalsize Experiments and Methods}
\subsection{\small AlN Wick Fabrication}
\par The wick fabrication does not require advanced chemical vapor deposition or single-crystal growth: a porous drop-cast hydrophilic wick was fabricated on a sheet of aluminum via a suspended slurry of 0.9 wt\% AlN (10,000 ppm suspended in DI water). The wicks were robust with repeatable evaporation data over multiple trials with the same wick when the fill factor was 40-60\% and the slab thickness was $\sim$200 $\mu$m. A slurry of AlN powder (10-$\mu$m diameter, Sigma Aldrich CAS 24304-00-5) with deionized (DI) water was drop-cast onto a sheet of aluminum on an open-air hot plate (300-400 $^\circ$C). The first layer was deposited at 300 $^\circ$C to ensure adhesion, and each subsequent layer was deposited at 400 $^\circ$C. The slurry was continuously shaken manually during the drop-casting procedure. Each layer was sequentially dried and returned to the hotplate temperature before the next layer was deposited. The thickness of the slab was controlled with the number of evaporation/deposition cycles, and the porosity was varied with the hot plate temperature and the slurry wt\% of AlN. The samples reused in experiments do not exhibit visible changes after reuse and show potential for salt-fouling resistance [See supporting information Sec. S5.2].

A thicker AlN slab would principally increase the optical path length, and correspondingly increase light-matter interactions. However, because of the drop-casting procedure, any irregularities in the evaporation are more pronounced when samples are thicker than 200 $\mu$m and reduce repeatability of experiments. Similarly, although a lower fill factor may be ideal\cite{Figliuzzi2013}, below $\sim$40\%, the AlN samples are fragile and disintegrate easily. Above a fill factor of $\sim$60\%, the samples do not saturate with water during experiments due to their restricted capillary size.

\subsection{\small AlN Slab Characterization}
\subsubsection{\footnotesize Thickness and Porosity}
The thickness of the slab was confirmed with scanning electron microscopy (Nova NanoSEM 450) and the fill factor was confirmed with a gravimetric method, i.e., measurement of the water-uptake mass with a fully saturated wick, using the bulk AlN density and measured mass \cite{Flint2018, Rao2022}. Consistent results were obtained when the fill factor was 40-60\% and the slab thickness was $\sim$200 $\mu$m.

\subsubsection{\footnotesize Chemical Composition}
\par Before and after fabrication, the sample composition was characterized via FTIR and XRD. After the wick fabrication, the chemical composition of AlN exhibited trace amounts of AlOH and Al$_{\rm 2}$O$_3$ [Fig. \ref{aln_props}]. The bulk properties appeared largely unchanged while the formation of micro-sheets of Al(OH)$_{\rm 3}$ was observed in small isolated regions. The Al(OH)$_{\rm 3}$ formation was attributed to hydrolysis and desorption of AlOH radicals from the DI water solvent during sample fabrication \cite{Pokorn2022, Kocjan2018}. The regions of Al(OH)$_{\rm 3}$ were small and isolated, since the aqueous AlN slurry was in minimal contact with DI water during fabrication. These AlOH and Al$_{\rm 2}$O$_{\rm 3}$ regions were expected to bridge the AlN in the wick without high-temperature sintering.


\subsubsection{\footnotesize Optical Characterization}\label{int-sphere}
\par The reflectance of the AlN slab was characterized with an integrating sphere (Newport 7540) with a photodiode and lock-in amplifier (Thorlabs PDA36A2 and Stanford SR830). The AlN slab reflected 28\%, 27\%, and 34\% of 405, 633, and 940-nm laser light, respectively. Significant waveguiding and diffraction was observed. The cone angle of diffraction scaled proportional to the wavelength of light on powder samples sandwiched between glass slides. 

\subsection{\small Experimental Procedure}
\par
\begin{wrapfigure}[27]{l}{0.4\textwidth}
    \begin{center}
    \includegraphics[width=0.4\textwidth]{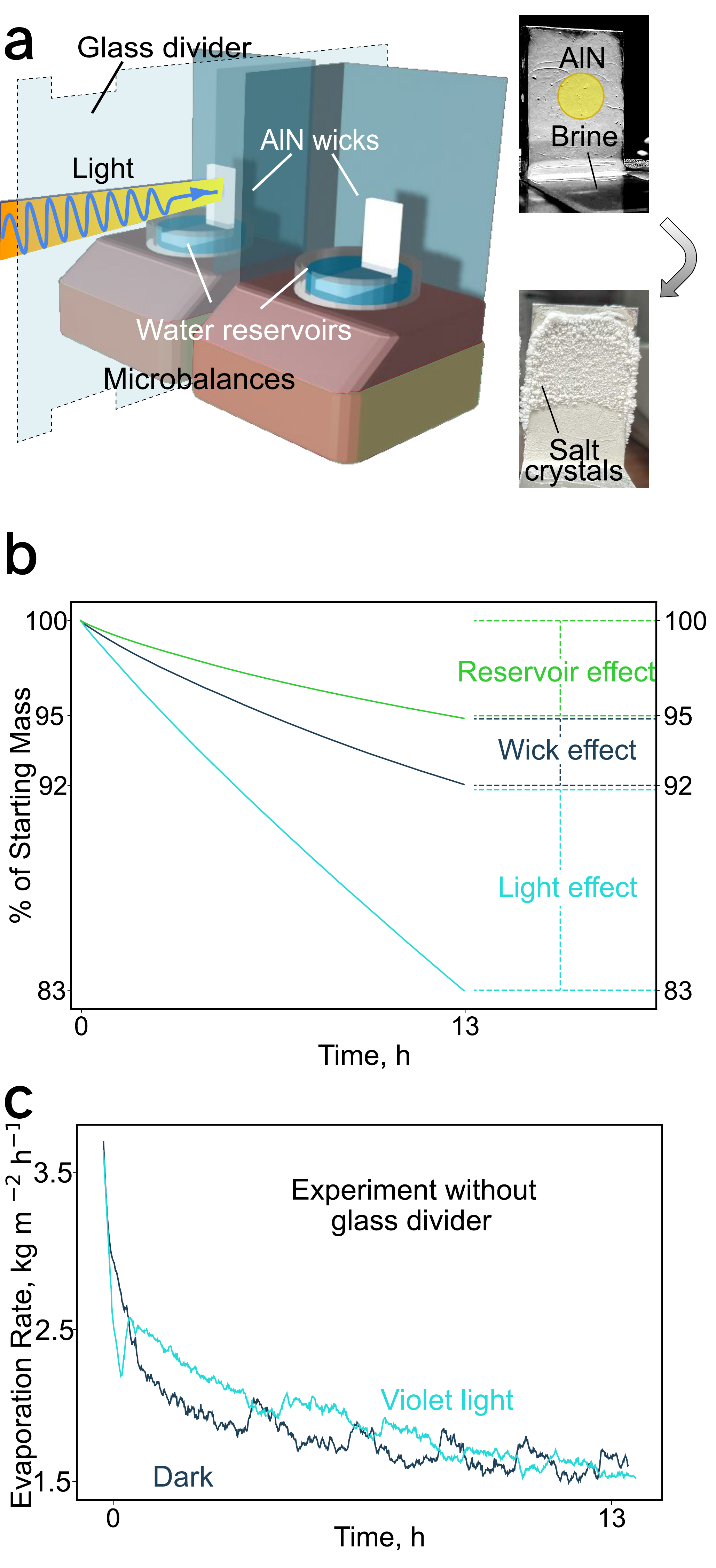}
    \caption{Experimental setup and representative raw data. (a) Schematic of the dual-balance self-referencing experimental setup. A glass divider with gaps reduces air flow between the balances. Inset sample photos: during water uptake showing light spot size (top) and salt nucleation (bottom). (b) Representative experimental data showing the relative mass lost due to the open reservoir, wick, and light. (c) Data showing cross-talk between samples with no glass divider.}
    \label{setup-raw-data}
    \end{center}
\end{wrapfigure}

 The effect of light was studied with a self-referencing system [Fig. \ref{setup-raw-data}(a)] that involved two microbalances (OHaus EX125) inside of a glovebox. The AlN wick samples [Fig. \ref{setup-raw-data}(a), photos] were placed in a water
\newpage

\noindent reservoir on both microbalances and an LED illuminates one sample. The mass data was captured with automated software. It was assumed that the mass lost from the evaporation at each setup occurred under similar environmental conditions because the humidity is similar on both sides of the glovebox [See Supporting Information, Fig. S3].

\par To characterize the system and separate the effect of the sample from the effect of light the background evaporation rate was measured (i.e., the mass change with no sample and no light), which quantified the increase in the evaporation from the AlN wick alone [Fig. \ref{setup-raw-data}(b)]. Subsequently, the measurement was

\noindent performed where one the AlN slabs was illuminated with LED light with a $\sim$1-cm$^2$ spot-size area and the other sample was the dark-field reference and not illuminated. 

\par For each set of data, the same wicks were reused as either the reference (un-illuminated) or measurement (illuminated). Low-intensity and high-intensity experiments were performed with 40 mW cm$^{-2}$ and 130 mW cm$^{-2}$, respectively. The power was measured with a photodetector and a power meter (Thorlabs S120C and PM100D) and the LED outputs were similar for different wavelengths of light at the sample [See Sec. S2 of the Supporting Information].

\par Air flow between the two balances was blocked with a glass divider. Figure \ref{setup-raw-data}(c) showed experimental data when no glass divider was present. Importantly, the divider did not fully partition the glovebox, and thus each sample equilibrates to similar climate conditions. Experiments lasting 800 minutes were conducted, during which time the air pressure, humidity, and temperature at a central location inside the glovebox were measured with a digital hygrometer (Fisher Scientific Traceable 6453). This dual-balance reference system minimized cross-talk between the dark reference and illuminated sample and produced similar environmental conditions over time to isolate the light-induced effect. The humidity fluctuated to a larger degree on the light-driven sample side of the chamber, but both sides were within 10\% relative humidity of each other. Farther from the sample, the humidity settled at ~65-70\% after 3-4 hours and remained relatively flat afterwards [See Sec. S2 of the Supporting Information].

\par Experiments measuring the evaporation were performed with high-intensity, high-salinity conditions with 385-nm, 625-nm, and 940-nm LED light [Fig. \ref{data-v-wavelength}(a), (c)]. The LEDs wavelengths are chosen to cover the range of wavelengths from near UV to NIR light and because similar characterization lasers with wavelengths of 405, 633, and 940 nm are available. Similar-wavelength lasers offer precise, non-destructive characterization of e.g. the sample’s reflectance and over long optical path lengths. While deep-UV illumination is expected to yield higher evaporation efficiencies than violet light, deep-UV LEDs with sufficient power and similar-wavelength laser diodes were not available at the time of this experiment. For 385-nm violet light, a matrix of experiments with three light intensities (off, low, and high) and three salinity concentrations (0, 9, and 25 wt\%) [Fig. \ref{data-v-salinity}] was also performed. These concentrations were significantly higher than even ocean water, which was only 3-4 wt\% salt, but are comparable to industrial runoff \cite{Bello2021, Panagopoulos2021}. 

\subsection{\small Efficiency Calculations} \label{EERCalc}
\par An above-unit energy conversion efficiency of 120$\pm$30\% was calculated for AlN with 385-nm violet light, and only 17$\pm$5\% and 10 $\pm$9\% for 625-nm orange and 940-nm IR LED light, respectively [Sec. S3 in Supporting Information]. In prior work, we established a model that accounts for the thickness of the sample [Ref. 47] and scattering angle. Here, we opt for a less granular analysis that is simply based on the absorbance of the sample. For more details on the calculation, see Sec. \ref{EERCalc}. For information on the reflectance measurement, see Sec. S1 in the Supporting Information. The glovebox ambient conditions were measured to be normal temperature and pressure (NTP, 20 $^\circ$C, 1 atm pressure), deviating $\pm$ 0.02 atm and $\pm$2 $^\circ C$. With sets of experiments that reused samples in different combinations of reference and measurement, the difference between the water evaporated from the sample with and without light was compared [Fig. \ref{setup-raw-data}(b)]. 

The wavelength-dependent energy conversion efficiencies were calculated with the average dark-field normalized mass lost, $\Delta m$, where the thermal heat needed to evaporate $\Delta m$ from the reservoir was 
\begin{equation}
Q_{\rm evap} = \Delta m (c \Delta T + L)
\label{eqn-q}
\end{equation}
where $c$ and $L$ was the salinity-dependent specific heat capacity and latent heat of vaporization for the reservoir salinity, and $\Delta T$ was the difference between the brine's elevated boiling point and NTP (84 $^\circ$C). The values of $c$ and $L$ were extrapolated from the published experimental trend \cite{Sharqawy2010, Generous2020}. For 25\% brine, the specific heat and heat of vaporization used were $c$ = 3.0 J g$^{-1}$ $^\circ$C$^{-1}$ and $L$ = 1.72 $\times$ 10$^3$ J g$^{-1}$, respectively [Sec. S3 in Supporting Information]. 

\par To isolate the evaporation that occurred from the light absorption in the wick, the difference in the illuminated and non-illuminated evaporation rates were compared. The efficiency of the light-enhanced evaporation was associated with the percentage of the absorbed light, which was either absorbed by the AlN or transmitted through the AlN and subsequently absorbed by the Al substrate [Fig. \ref{data-v-wavelength}(b)]. Both of these components produced heat in the reservoir, which contributed to the calculated evaporation efficiency, $\eta$,

\begin{equation}
    \eta = \dfrac{Q_{\rm effective}}{Q_{\rm absorbed}},
    \label{eqn-eta}
\end{equation}

\noindent where $Q_{\rm effective} = Q_{\rm evap, light}-Q_{\rm evap, dark}$ was the light energy used by the entire system, and $Q_{\rm absorbed} = q_{\rm in}(1-R_{\rm sample})At$ was the total light energy absorbed by the sample over the course of an experiment, where $q_{\rm in}$ was the measured flux of the LED, $R_{\rm sample}$ was the sample's measured reflectance, $A$ was the area of the sample which was illuminated, and $t$ was the time of an experiment [Secs. S1 and S2 in the Supporting Information]. $Q_{\rm effective}$ accounts for the enhanced evaporation that occurs by increasing the specific surface area of the air-water interface where $Q_{\rm evap, light}$ and $Q_{\rm evap, dark}$ correspond with the measurement and reference experiments [Eq. \ref{eqn-q}].
Although the energy delivered to the sample was $\sim$6.2 kJ over an entire experiment, the evaporation rates used to calculate $\eta$ in Eq. \ref{eqn-eta} were extrapolated from the stable evaporation conditions with the energy and mass lost between $400$ and $800$ minutes in the experiment. In this case, $q_{\rm in}$ was $\sim$3.1 kJ, and $Q_{\rm absorbed}$ is 2.1 - 2.3 kJ, depending on the wavelength of light.

\section{\normalsize Results and Discussion}
\subsection{Results}

\begin{figure}[t!]
    \centering
    \includegraphics[width=\textwidth]{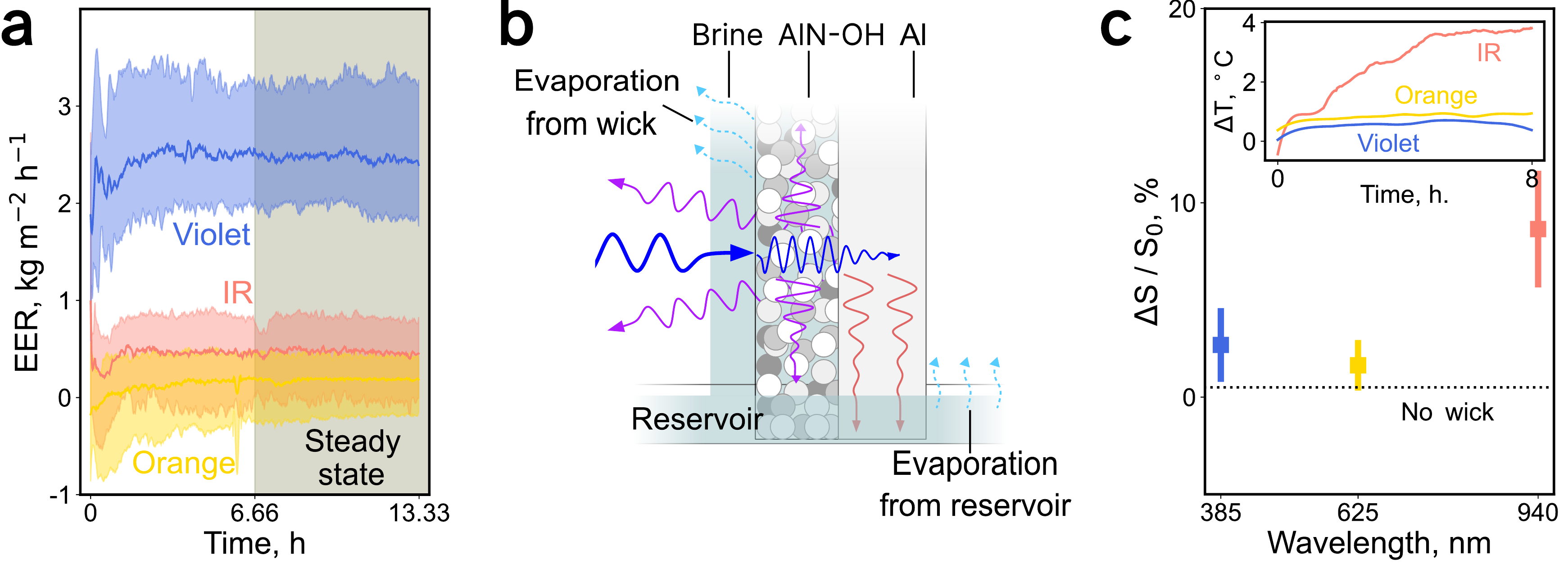}
    \caption{Spectral dependence of the light-enhanced evaporation rate (EER) with a 25-wt\% salinity reservoir. (a) EER vs. time for experiments with
 385-nm (violet), 625-nm (orange) and 940-nm (IR) light. (b) Schematic showing energy paths in the porous AlN sample. Incident light energy ($Q_{\rm in}$) is reflected, waveguided, and transferred  to the reservoir and associated with thermal loss ($Q_{\rm res}$). (c) Percent salinity change in concentration $\Delta S/S_{\rm 0}$ in the reservoir. Inset plot: temperature increase $\Delta T$ vs. time. The illumination intensity is 130 mW cm$^{-2}$.}
    \label{data-v-wavelength}
\end{figure}

\par Initially, the presence of a dry AlN wick draws an outsized volume of water, which increases the mass-change measurement. The wick becomes saturated within the first three minutes of an experiment, at which point the mass-change measurements are largely attributed to evaporation. After the wick saturates with water, enhanced evaporation occurs as the humidity increases inside the glovebox. The evaporation dynamics stabilize within the first four hours, at which point the relative humidity inside the glovebox is $\sim$70\%. We refer to this equilibrated evaporation regime of the experiment as the steady-state [See gray-shaded area in Fig. \ref{data-v-wavelength}(a)]. The dark-field control measurement quantifies the effect of the dry wick without light. A dry wick without light will evaporate 50\% more water than the reservoir without a wick. In the case with 25 wt\% salt water with 130-mW cm$^{-2}$ intensity violet light, the evaporation doubles further [Fig. \ref{setup-raw-data}(b)].

\par
\begin{figure}
    \centering
    \includegraphics[width=0.8\textwidth]{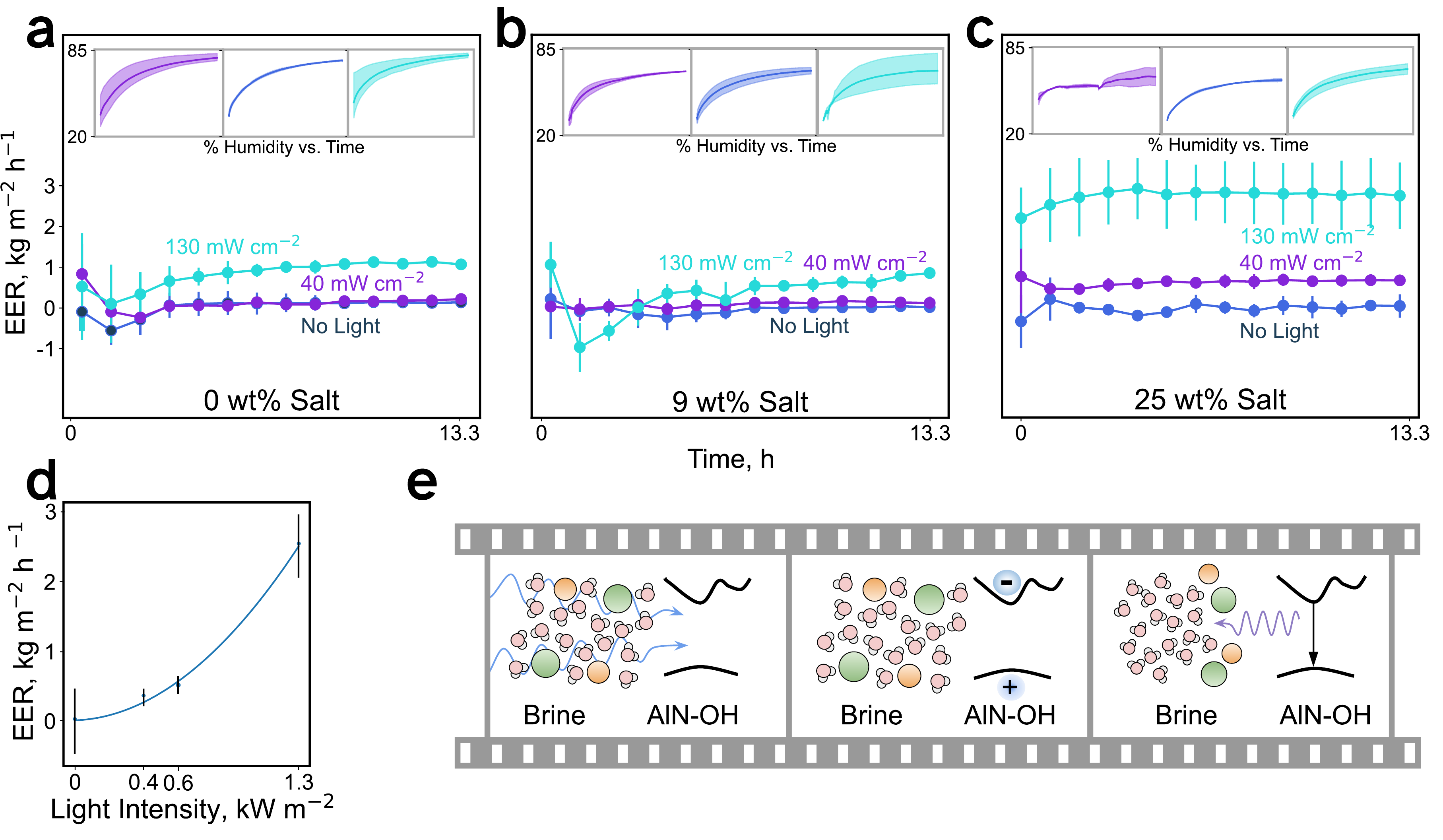}
    \caption{Evaporation trends and mechanism with violet light. (a)-(c) EER as the salinity increases from zero to 25 wt\% and the light intensity increases from 0 to 130 mW cm$^{-2}$. (d) EER for high-salinity violet light experiments plot as a function of light intensity showing a quadratic dependence. (e) Schematic showing propose upconversion in bulk AlN and at -OH interface that targets brine evaporation and facilitates salt nucleation.}
    \label{data-v-salinity}
\end{figure}

 We identify thermal leakage or $Q_{\rm res}$ from the measured evaporation rates with zero wt\% salt [Fig. \ref{data-v-salinity}(a)], which lowers the calculated efficiency. The evaporation enhancement rate (EER) is the discrete time derivative of the mass flux with reference to the dark-field experiment. We observe a nonzero EER of $1.0 - 1.1$ kg m$^{-2}$ h$^{-1}$ for light intensities of 130 mW cm$^{-2}$. This indicates that our experiments are 30-50\% photothermal, i.e., there is heat produced and thermal leakage between the slab, substrate, and reservoir. In other words, we expect an EER of zero when no energy is transferred from the wick to the reservoir since water is transparent to visible light. The magnitude of $Q_{\rm res}$ is half that of $Q_{\rm effective} = Q_{\rm light}- Q_{\rm dark}$ for high salinity brine and violet light [Fig. \ref{setup-raw-data}(b)].

\par Reservoir temperature measurements and commensurate changes of the reservoir salinity confirm nonzero EER with light and significant $Q_{\rm res}$. Less energy is required to evaporate salt water compared to fresh water even though the addition of salt produced an elevated boiling-point, since $c$ and $L$ decrease. The photothermal limit is defined as the maximum amount of evaporation achievable with one-sun intensity assuming 100\% light absorption and thermal heat conversion for the reservoir salinity. For 25 wt\% salt water, the photothermal limit is 1.82 kg m$^{-2}$ h$^{-1}$, calculated with Eq. \ref{eqn-q}.

We associate changes in the reservoir salinity with absorption of light in Al and the corresponding thermal changes in the reservoir. The equilibrium temperature increases 0.5 $^\circ$C for violet and orange light and 3.5 $^\circ$C for IR light with 130 mW cm$^{-2}$ intensities [Fig. \ref{data-v-wavelength}(c)]. At the same time, the changes in the reservoir salinity associated with the illuminating light are $\sim$3\% for violet and orange light and $\sim$9\% for IR light.  With violet and orange light, the evaporation and temperature/humidity dynamics stabilize within four hours. The temperature/humidity with IR light take longer to stabilize, which we associate with a higher $Q_{\rm res}$ for IR light [Fig. \ref{data-v-wavelength}(c)].

\par In spite of thermal losses, with steady-state conditions and 70\% humidity, the calculated efficiency $\eta$ is between 90\% and 150\% for violet light and only 20\% for both orange and IR light. The raw change in mass is nearly double with violet light compared to orange and IR light; when we subtract the dark-field mass change from each wavelength measurement, the EER with violet-light is significantly higher [Fig. \ref{data-v-wavelength}(a)]. Details of the calculations are in Sec. S3 of the Supporting Information, and a summary of the calculated, measured, and derived values is in Sec. S4 of the Supporting Information. If we include the steady-state wick evaporation, the efficiency $\eta$ also increases and is 150-270\% for violet light, and 70-100\% for orange and IR lights. Notably, our $\eta$ calculations are performed with steady-state evaporation rates measured under steady-state conditions; in an open-air environment, the evaporation rates associated with both light and the wick would increase further.

\par We present a lower-bound efficiency, which does not account the dynamic salinity or thermal conduction and losses. We use a fixed value for the specific heat capacity, absorption, and latent heat of vaporization. This assumption is reasonable for violet and orange-light experiments where the salinity does not change significantly over the course of the experiment [Fig. \ref{data-v-wavelength}(c)] and the temperature stabilizes within a couple of hours. We ignore the absorption of the brine film at the AlN interface because violet, orange, and IR light are not readily absorbed by the brine at these wavelengths [Fig. \ref{motivation}(a)]; the changes in light absorption from the brine, as well as the nucleated salt crystals on the AlN are also considered negligible \cite{Tong2020, Miyata1968}.

\subsection{\normalsize Discussion} \label{disc}
\subsubsection{\normalsize Salinity Dependence and Fluid Dynamics}

\par The ultimate EER is limited by the vapor saturation of the ambient air\cite{Gupta2014}; once the ambient conditions stabilize, the evaporation rate stabilizes. For each experiment, we observe a higher short-term EER during the first five minutes to an hour, and a lower EER as the environment stabilizes, which occurs generally around 70\%-humidity [Fig. \ref{data-v-salinity}(a-c)]. The EER dynamics are consistent but vary depending on the combination of samples used: differences in the wick surface smoothness are expected to cause wick-specific trends. Issues with clogged salt pores and asymmetry in the test chamber may contribute to irregularities, particularly with the high, 25 wt\%-salinity brine. Nevertheless, the light enhanced effect with violet light is three times larger than the variance of each measurement when considering equilibrium conditions with 70\% humidity. 

The competition between advection and diffusion dynamics can be inferred from the EER salinity-dependent inflection point i.e., the low 9 wt\%-salinity EER compared to both zero wt\% and 25-wt\%-salinity EERs [Fig. \ref{data-v-salinity}(a-c)]; advection is dominant at low concentrations when the concentration gradient is low while diffusion is dominant at high concentrations when the salinity gradient is high \cite{NorouziRad2012}. An opposite trend is derived in the Pitzer model\cite{Pitzer1984}: a lower evaporation rate is expected for fresh water than with brine, however in our system, there are additional inhomogeneities that explain this trend. For example, advection may be low with violet light since light is more strongly absorbed by brine than by fresh water. These trends make it difficult to study the optical effect alone but indicate important opportunities to control, design, and amplify the fluid dynamics and enhance evaporation further.

In general, we observe a higher EER with increased salinity. There are several explanations for this trend. Firstly, the evaporation energy for salt water is lower than for pure water, so photothermal effects enhance the evaporation to a higher degree with increasing salinity. Secondly, the presence of salt crystals and surface roughness may serve as the nucleation sites for enhanced mineralization that would increase at higher concentrations. Our results suggest that it is not a thermal response that dominates the EER trends. Both violet and orange light reflect similar light energies, have similar thermal losses, but the EER from violet light is $\sim10\times$ higher, leading to $\sim6\times$-higher light conversion efficiencies.

\subsubsection{\normalsize Spectral Dependence and Bounds for Efficiency}
The spectral dependence of the efficiency $\eta$ is remarkable: the EER with 385-nm violet light is almost 10$\times$ higher than with 625-nm orange light, despite similar estimated AlN absorption. Both wavelengths exhibit similar reflectivities and have similar thermal losses: the absorption coefficient is higher for violet light but orange light has a longer effective optical path length, i.e., waveguided in AlN \cite{Singh2024}. Heating may assist in the evaporation, but comparative experiments of violet and IR light indicate that heat is not the dominant effect with violet light. In fact, the reservoir is heated to a greater degree with IR light [Fig. \ref{data-v-wavelength}(c)], which leads to efficiencies that are $\sim 10\times$ higher with violet light.

As previously mentioned, our light-induced efficiency calculations are conservative since they assume constant values for the brine-dependent heat capacity and latent heat of vaporization even though the salinity of the reservoir increases. Higher-salinity brines evaporate more readily and we do not account for this salinity increase in our calculations. We claim a lower-bound efficiency that incorporates sample absorption and dark-field referenced evaporation measurements.  While we can infer significant energy transfer to the reservoir and the aluminum substrate by changes in temperature and salinity, we only use the quantified sample absorption from the integrating sphere measurements for the efficiency calculations. The measured efficiency is a lower bound, where the denominator of the efficiency calculation is as large as possible.

On the other hand, chamber fluid dynamics accompany the wick evaporation and point to the intrinsic challenge of isolating the EER. In fact, without the glass divider, we observe crosstalk between the light and dark microbalance measurements [Fig. \ref{setup-raw-data}(c)]; the presence of the glass divider does not fully eliminate environmental gradients and other chamber barriers such as mesh screens may improve future designs \cite{Mehta1985, Zhu2017b}. The presence of convective dynamics may be related to higher reported EERs accompanying higher optical intensities in other reported photothermal systems, i.e., five to ten-suns ($\sim4-8\times$ higher than our values) \cite{Zhu2017, Zhu2017a, Zhu2018, Zhang2020a, Xu2019, Wang2020, Song2020, Qiao2019, Ni2018, Ma2019, Kuang2019, Liang2019, He2019, Ma2023} especially since differences as much as 20\% evaporation efficiencies have been observed between one-sun and ten-sun illuminations \cite{Zhu2017, Zhu2017a, Kuang2019, Liang2019, He2019, Chen2017}. Our closed-system analysis may not fully eliminate the effect of convection, however the spectral trend associated with different wavelengths of light is clearly observed, which points to a non-photothermal evaporation pathway. 

\subsubsection{\normalsize Analysis of a Non-photothermal Mechanisms}\label{nonphoto}
\indent Deep-UV photons are selectively absorbed by salt-water bonds and this reveals a spectrally selective path to non-photothermal desalination: if lower-energy visible light produces high-energy deep-UV photons in salt water, the salt-water bonds will be selectively targeted with minimal heating of bulk water. The quadratic dependence of the EER with violet light intensity supports the hypothesis that EER is influenced by a photon upconversion or second- harmonic processes, i.e., two photons combine to create charge transfer with double the energy  [Fig. \ref{data-v-salinity}(d)]. Second-harmonic generation, where deep UV 200-nm photons are generated from violet 400-nm photons, would produce a similar nonlinear trend [Sec. S1 in Supporting Information]. We illustrate the nonlinear photon upconversion process in Fig. \ref{data-v-salinity}(e). 

While our experiments do not provide experimental evidence of the mechanisms at work, we hypothesize that deep-UV photon upconversion is supported by processes associated with both AlN in the material bulk and at its surface. In the bulk, second harmonic generation and {phonon-assisted nonlinearities are} known to be outsized with non-centrosymmetric wurtzite-crystal AlN \cite{Hughes1997, Aguilar2019, Liu:23}. At the interface, the FTIR spectra shows strong hydroxyl-group -OH enhancement [Fig. \ref{aln_props}(a)]. The role of surface water is intriguing; the ``facile exchange of electronic and vibrational energy with the majority of the excitation energy stored as O–H vibrational energy in a 14–15 \AA{}\ layer of water'' is previously reported by Neuweiler and Gafney with glass \cite{Neuweiler2019}. It has been shown that a surface -OH termination plays a crucial role in increasing hydrophilicity in metalloids \cite{Chen2018}. The clear presence of Al-OH groups in the sample AlN FTIR spectra may drive energy transfer between the Al and AlN and the surrounding salt water. We observe that the presence of Al-OH in the FTIR data correlates to the rate of capillary rise or hydrophilicity in the wick, in preliminary experiments. The bulk absorption, harmonic generation, and interfacial energy transfer processes associated with surface water and the bulk bandgap would serve as spectral selective drivers for the non-photothermal upconversion and salt-water EER.

In a similar manner with their 2023 PNAS paper, ``Plausible photomolecular effect leading to water evaporation exceeding the thermal limit'', Yaodong Tu and Gang Chen proposed a mechanism for non-photothermal light-enhanced evaporation \cite{Tu2023, Lv2024}. Central to their observation is that for pure water, the inter-molecular binding forces are an order of magnitude lower than the energy of visible light, and therefore light carries enough energy to drive rapid evaporation. However, because of the limited degrees of movement of water molecules in bulk, as well as the bulk dipole charge-alignment, the light energy is dispersed within the fluid and cannot be utilized effectively. Immediately at the air-water interface, however, there exists a monolayer of high-energy water molecules suspended, which are free to align without the net-zero charge alignment restriction that is ordinarily assumed for bulk water. By confining their observations to this monolayer, where water molecules can evaporate more freely than in bulk, the authors show that light has an extraordinarily high ability to cleave off water molecules with remarkable energy efficiency. They support this claim in several ways, with myriad evidence, and the linchpin of their observations stems from the binding energy of pure water and of visible light. Like Tu and Chen, Neuweiler and Gafney  hypothesize that surface -OH facilitates energy exchange and charge transfer with remarkably low light energy compared to the electronic bandgap and even observe Fibonacci harmonics (referred to in the paper as ``overtones'' \cite{Neuweiler2019}). In our work, we provide experimental observations of this non-photothermal effect via both bulk and interfacial effects. Future spectroscopic analysis of the role of surface -OH groups facilitating EER could further advance such claims of non-photothermal evaporative interactions associated with surface water.

Water-related experiments are often challenging and the research can be speculative. Here we present efficiencies based on careful sample optical absorption and dark-field measurements.  Thermal effects are present as evidenced by the salinity and rise in temperature and result in lowered efficiency.  We believe the primary source of error would reside in the presence of convective rolls induced by light.

\section{\normalsize Conclusions}

In conclusion, we demonstrate the light-induced and enhanced saltwater evaporation with a hard, white, ceramic wick where the salt nucleates onto the wick, providing a path for ZLD solar desalination. Through careful experiments, we isolate and study the spectral selection of AlN wicks, wherein the energy conversion efficiency increases with blue wavelengths and higher-salinity brines. We account for “ambient thermal inputs”-- dynamic changes in the environment-- with a dark-field measurement. Our results point to energy conversion efficiencies above the photothermal limit with violet light and a spectrally selective path to target saltwater evaporation with minimal heating. We hypothesize that a combination of bulk and surface hydroxyl-group interactions offer upconversion to drive deep-UV interactions and break salt-water bonds. Notably, our experiments show high efficiency with monochromatic light, which is not representative of the solar spectrum. At the same time, the practically-unity wall-plug-in efficiency of blue LEDs \cite{Kuritzky2018} could mean that highly-efficient LED-pumped desalination systems are feasible. Our results point to important considerations for the spectral design of solar desalination materials and interfaces. 

\section{\normalsize Supporting Information}
Reflectance measurements, notes about the experimental setup and data normalization, efficiency calculation details, tables of: material coefficients, summary-at-a-glance of evaporation rates, and literature comparison.

\section{\normalsize Acknowledgements}
Authors gratefully acknowledge funding from NSF DMR 1921034. Navindra is supported by a Department of Education Graduate Assistance in Areas of National Need (GAANN) Fellowship.  Other support comes from Winston Chung Global Energy Center (WCGEC) Seed Research Grant and dedication of students, as well as helpful conversations with Alex Greaney and Bhargav Rallabandi. The authors also gratefully acknowledge the time and attention from reviewers, whose comments improve the manuscript.

The following files are available free of charge.
\begin{itemize}
  \item Supporting Information pdf
\end{itemize}

\bibliography{refs.bib}
\end{document}


\runningpagewiselinenumbers
\maketitle
\linenumbers


\pagebreak
\section*{Sec. S1: Reflectance Measurement}
\begin{wrapfigure}[9]{r}{0.6\textwidth}
    \begin{center}
    \vspace{-15mm}
    \includegraphics[width=0.6\textwidth]{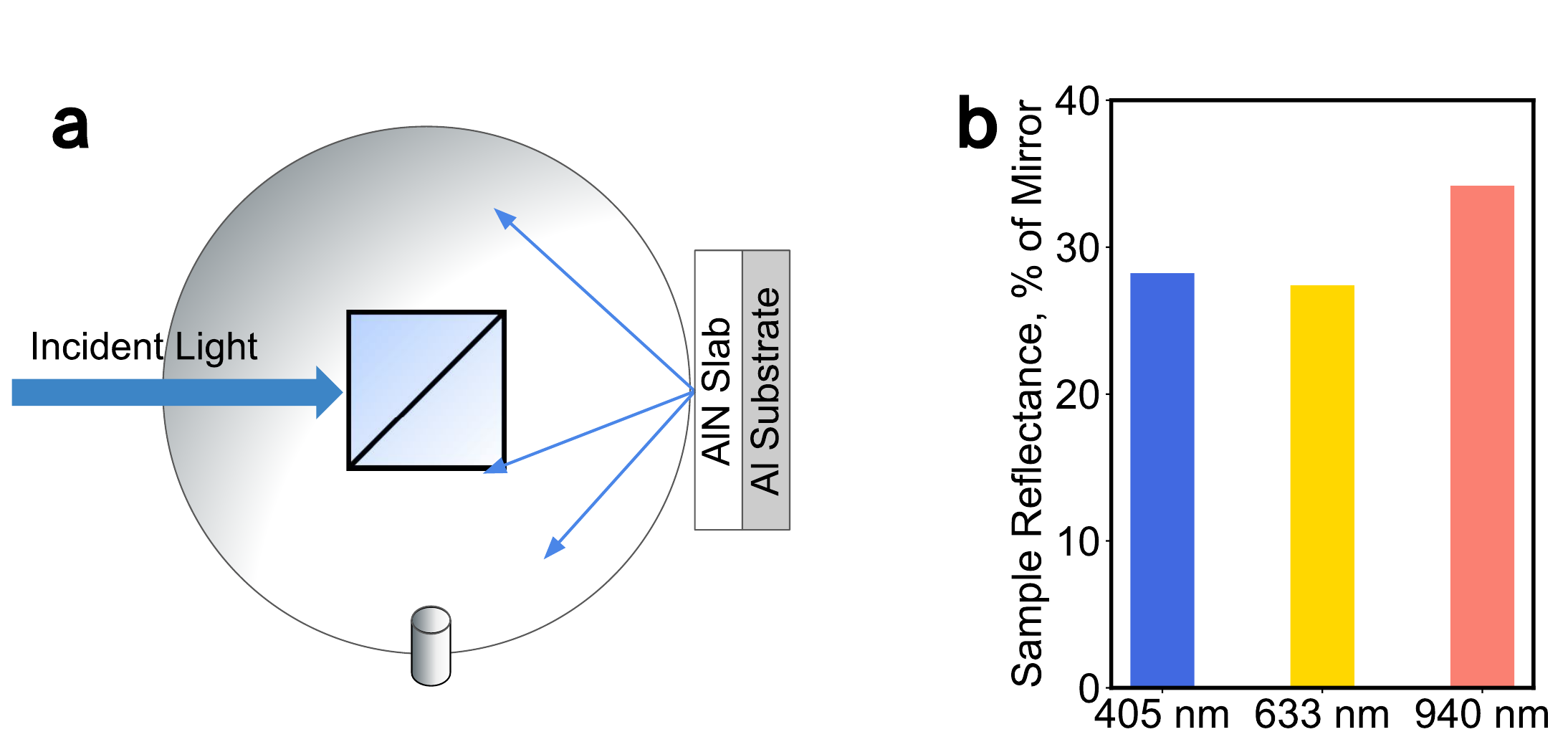}
    \end{center}
    \caption{(a) Diagram of integrating sphere. (b) Polished mirror and sample reflectance results. We conclude that the AlN slab reflects 27-28\% at both violet $\lambda = 405$-nm and orange $\lambda = 633$-nm wavelengths.}
    \label{reflectance_msmt}
\end{wrapfigure}
We measure the sample reflectance with an integrating sphere (Newport 7540) and lock-in amplifier (Stanford SR830) at violet and orange wavelengths. The measurements are performed with lasers whose wavelengths are close to that of the LEDs used in the evaporation experiments. The laser is passed through an optical chopper, which is referenced with the lock-in amplifier. We calculate the sample reflectance, 

\begin{equation}
    R_{\rm sample} = \dfrac{R_{\rm AlN} - R_{\rm background}}{\frac{R_{\rm mirror, \,meas}}{R_{\rm mirror, \,spec}} - R_{\rm background}},
    \label{eqn-r_sample}
\end{equation}
where the measured and specification reflectance from a polished silver mirror (Thorlabs PF-10-03) are $R_{\rm mirror, \,meas}$ and $R_{\rm mirror, \,spec}$, respectively. This accounts for the integrating sphere coating. The background measurement, $R_{\rm background}$ is taken with the laser off and $R_{\rm AlN}$ is measured with the sample. The raw voltage measurements are in Table \ref{table-reflectance_msmt}. The result is that the AlN samples reflect similarly for both violet (28\%) and orange (27\%) light and slightly more for IR (34\%) light.

\begin{table*}[t]
\begin{center}
    \rowcolors{2}{gray!25}{white}
    \begin{tabular}{cccccc}
    \hline
    Wavelength, nm
    & $R_{\rm AlN}$, mV
    & $R_{\rm background}$, mV
    & $R_{\rm mirror, meas}$, mV
    & $R_{\rm mirror, spec}$
    & $R_{\rm sample}$, \% \\
    \hline
    405
    & 0.295
    & 0.014
    & 0.92
    & 0.91
    & 28 \\
    633
    & 0.1
    & 0.004
    & 0.35
    & 0.99
    & 27 \\
    940
    & 1.1
    & 0.014
    & 3.16
    & 0.99
    & 34
    \end{tabular}
    \caption{Integrating sphere voltage measurements for AlN sample and mirror reference}
    \label{table-reflectance_msmt}
\end{center}
\end{table*}

\pagebreak
\section*{Sec. S2: Experiments and Data Processing}

\subsection*{Sec. S2.1: Experimental Setup}
\begin{wrapfigure}[21]{l}{0.5\textwidth}
    \begin{center}
    \vspace{-15mm}
    \includegraphics[width=0.5\textwidth]{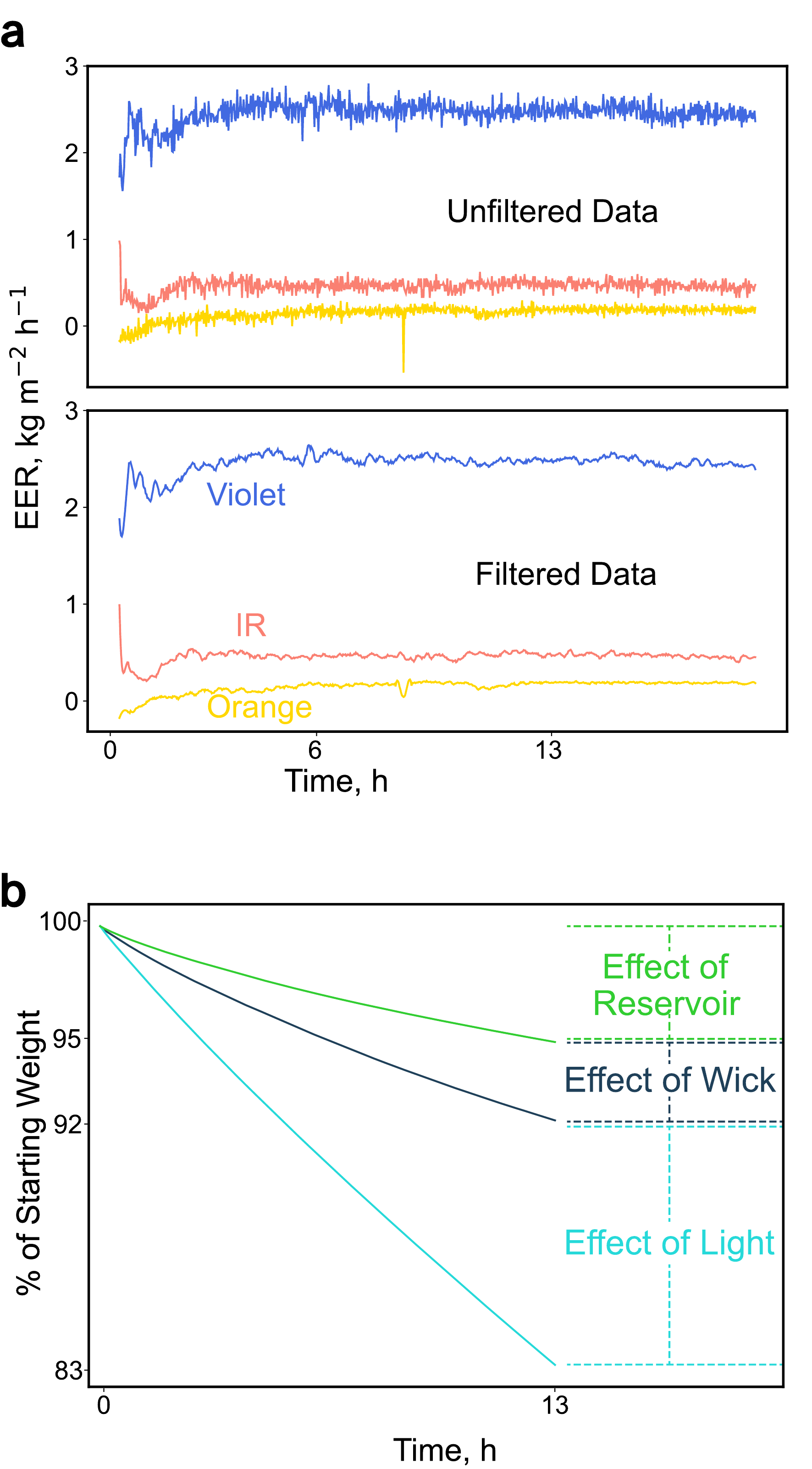}
    \end{center}
    \caption{(a) Unfiltered vs. Savitsky-Golay filtered normalized data. (b) Representative raw experimental data, showing the evaporation enhancement of the wick and light separately.}
    \label{fig-raw-data}
\end{wrapfigure}

We perform experiments inside of a sealed glovebox that contains two microbalances (OHaus Explorer EX125). Each microbalance measures the change in mass from a $\sim$40-mL reservoir of water. Additionally, a digital hygrometer (Fisher Scientific Traceable 6453) is placed at the center inside the glovebox to continually record the temperature, pressure, and humidity. A vertical, free-standing AlN wick is placed inside the reservoirs for experiments which are done with a sample. Alternative ceramics were considered as a reference, but due to inconsistent fabrication, the results are not presented here. The salinity of the reservoir is varied between 0 wt\% and 25\% during different experiments. One sample is illuminated with light from an LED (Thorlabs M385LP1, M625L4, or M940L3), which delivers 130 mW cm$^{-2}$ over 1-cm$^2$ area at the sample. Note that the LEDs employed have wavelengths of 385 nm, 625 nm, and 940 nm respectively and closely match the wavelengths of the low-power lasers used for optical reflectivity measurements [Sec. S1]. Between the two microbalances, a glass pane is placed to minimize convective cross-talk [Fig. 3 (c) in the main text]. There is a $\sim$2 cm$^{2}$ gap on the top and bottom of the glass pane. The glovebox allows for the reference and control measurements to be performed under similar environmental conditions, while isolating both setups from large fluctuations in humidity, temperature, and pressure. Without a divider, there is significant cross-talk between setups and without the wick, the evaporation is significantly reduced [Fig. \ref{fig-raw-data}].

\begin{wrapfigure}[11]{l}{0.5\textwidth}
    \begin{center}
    \vspace{-10mm}
    \includegraphics[width=0.5\textwidth]{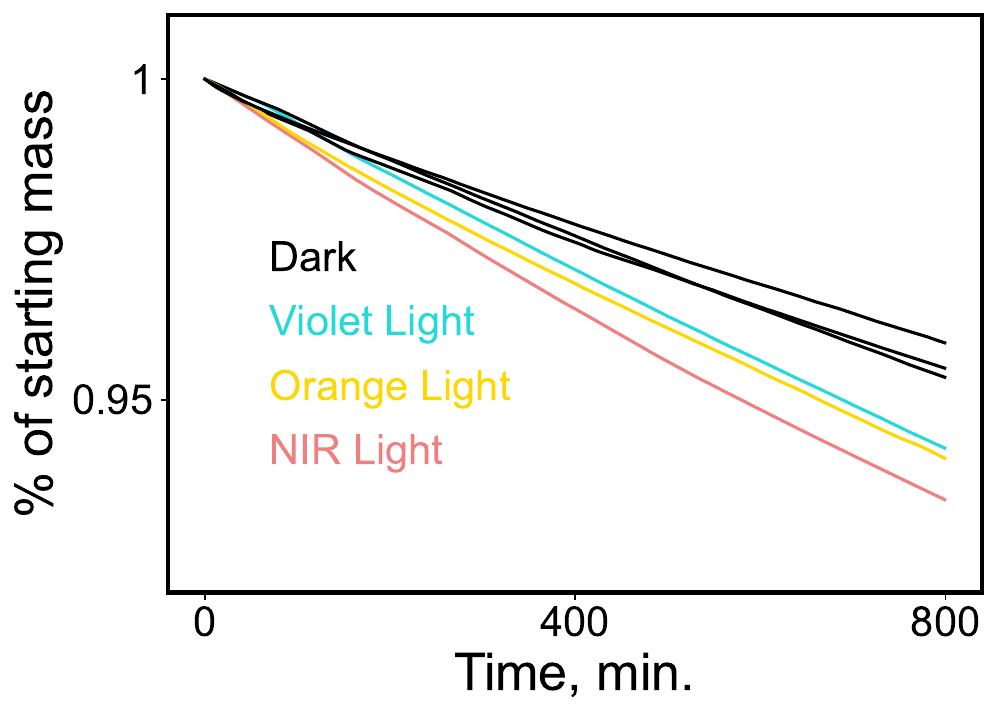}
    \end{center}
    \caption{Mass loss vs. time for direct illumination of 20 wt\% saltwater brine with violet, orange, and NIR light.}
    \label{fig-direct_water}
\end{wrapfigure}

Figure \ref{fig-direct_water} shows the direct evaporation of salt water when illuminated with 1-sun intensity light at 385, 625, and 940-nm light. The overlap between all three sets of experimental data show that salt water on its own does not exhibit spectral-selective evaporation enhancement in the near UV. We have conducted preliminary experiments on the light-induced saltwater evaporative cooling\cite{Singh2025} These trends indicate that blue light on AlN leads to enhanced cooling compared to saltwater or freshwater with orange and IR light.

\subsection*{Sec. S2.2: Data Normalization}

The raw output from our experimental setup is a 1-D array consisting of the instantaneous mass at the time of software output. To graph the data, we need to account for the polling interval of the microbalance, which we set to one minute. When we compute the discrete derivative [Fig. \ref{fig-raw-data}], the units are g min$^{-1}$. Owing to the spot size of our light, the actual units can be taken as g 0.8-cm$^{-2}$ min$^{-1}$. When we convert to the industry-standard kg m$^{-2}$ h$^{-1}$, the scaling factor that pops out is $\times$750. Only a fraction of prior research uses a sample in their dark-field reference. Accordingly, the data normalization procedure we adopt is to subtract the dark-field reference, compute the discrete derivative of the mass with respect to time, then multiply the resulting array by 750. In addition, data plots in the main text have been processed via a Savitsky-Golay filter from the scipy signal library, with a window size of 80 and a polynomial order of 4.

When calculating the efficiency of the light-accelerated evaporation, the area used is the spot size of the LED beam, equal to 0.8 cm$^{-2}$. This is common practice in experiments where only a portion of the sample is illuminated \cite{Zhang2023, Xu2020, Wang2020, Ni2016, Ni2018}.

\subsection*{Sec. S2.3: Nonlinear Evidence of Non-Photothermal Light-Induced Evaporation}

\begin{wrapfigure}[10]{l}{0.3\textwidth}
    \begin{center}
    \vspace{-10mm}
    \includegraphics[width=0.3\textwidth]{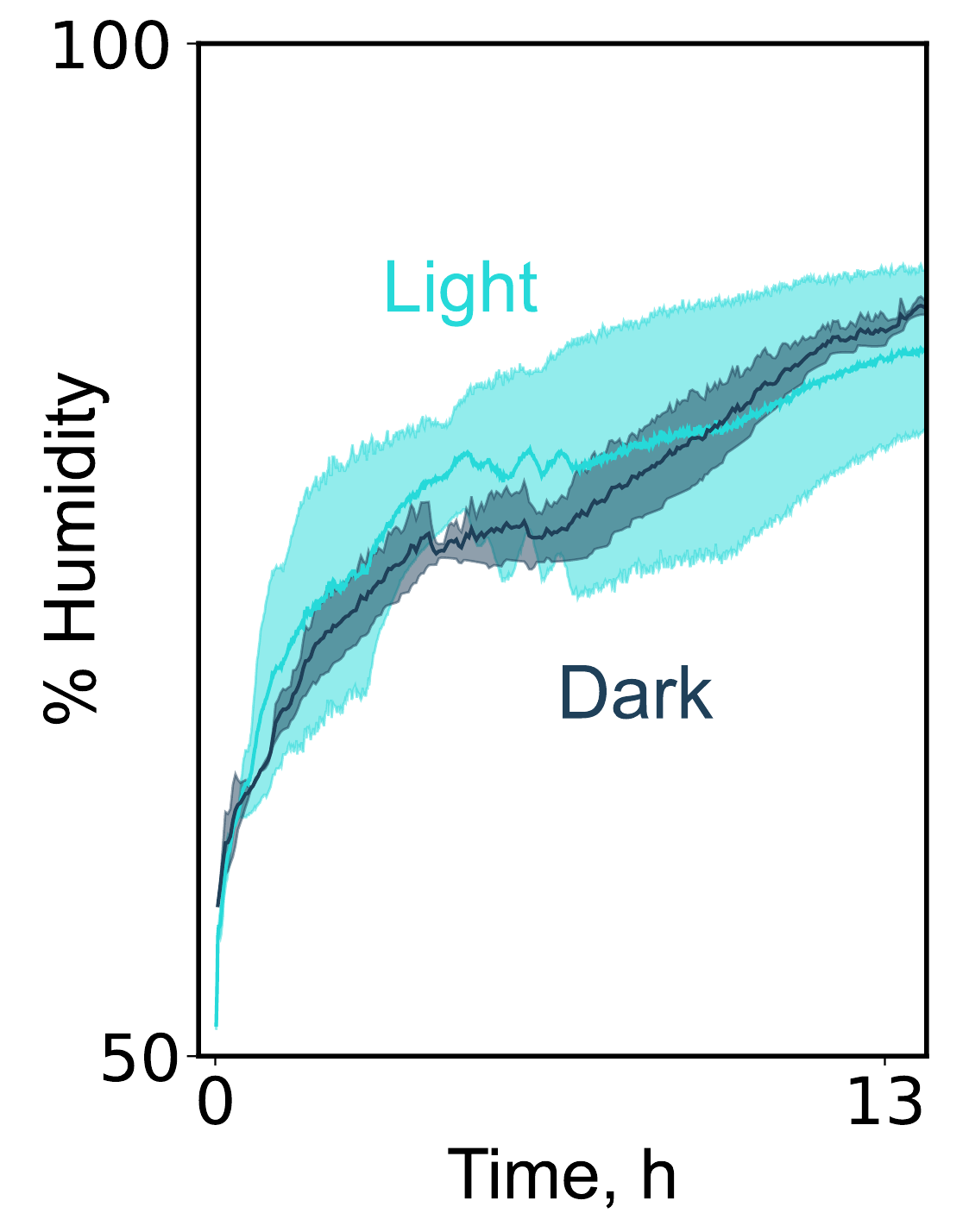}
    \end{center}
    \caption{Humidity data during repeat experiments, measured on both sides of the chamber, i.e. light and dark reference.}
    \label{fig-humidity}
\end{wrapfigure}

We estimate the contribution of the incident beam and its generated second harmonic by establishing a system of equations based on experimental and calculated results [Fig. 5(d) in the Main text]. The intensity of energy absorbed from the incident beam is proportional to the input intensity. This is in contrast with the power of a generated second-harmonic signal, which varies in proportion to the input intensity squared \cite{Agrawal2013}. We can leverage this fact to establish the following system of equations:

\begin{eqnarray}
    EER_{1} & = & a + bI_{1} + cI_{1}^2\\
    EER_{0.5} & = & a + bI_{0.5} + cI_{0.5}^2\\
    EER_{0.3} & = & a + bI_{0.3} + cI_{0.3}^2\\
    EER_{0} & = & a + bI_{0} + cI_{0}^2
\end{eqnarray}

\noindent where $EER_{n}$ is the evaporation enhancement rate at n-sun light intensity, $b$ and $c$ are the weighting terms that scale the evaporation rate linearly (associated with light absorption and thermal evaporation) and quadratically (associated with second harmonic) and $I_{n}$ is the absorbed light energy at n-sun intensity [Sec. S1 in Supporting Information]. Substituting experimental values of for the $EER_{n}$ [Fig. 5(d)], we arrive at non-zero solutions for $a$, $b$, and $c$. The quadratic curve of fit has a mean-squared error of $<$0.5\% for the average of our experimental data, and 2.1\% for the error-bars of repeat experiments.

\section*{Sec. S3: Light-induced Wick Evaporation and Efficiency Calculation}
\par The key equation for calculating the efficiency of our system is the phase-change equation:

\begin{equation}
        Q = m(c\Delta T + L)
        \label{eqn-phase_change}
\end{equation}

\noindent Here, $\Delta T$ is the difference between the water's initial temperature and its salinity-dependent boiling point, and $c$ and $L$ are the specific heat capacity and vaporization enthalpy, calculated using Eqs. \ref{eqn-L} and \ref{eqn-c_p}
\begin{equation}
    L = (a_{1} + a_{2}T + a_{3}(T-100)^2 + a_{4}T^3 + a_{5}T^4)(1 - \frac{S}{1000})
    \label{eqn-L}
\end{equation}

\begin{eqnarray}
    A & = & d_{1} + d_{2}T + d_{3}T \nonumber \\
    B & = & b_{1} + b_{2}T + b_{3}T \nonumber \\
    c(T) & = & c_{1} + c_{2}T + c_{3}T^{2} + c_{4}T^{3} + c_{5}T^{4} \nonumber \\
    c(T,S) & = & c(T) + AS + BS^{1.5}
    \label{eqn-c_p}
\end{eqnarray}

\begin{table*}[ht!]
\begin{center}
    \begin{tabular}{cc|cc|cc|cc}
    a$_{1}$ & 2.501$\times$10$^{6}$ &
    b$_{1}$ & 0.1770383 &
    c$_{1}$ & 4.2174$\times$ 10$^{3}$ &
    d$_{1}$ & -7.643575 \\
    a$_{2}$ & -2.369$\times$10$^{3}$ &
    b$_{2}$ & -4.07718 $\times$ 10$^{-3}$ &
    c$_{2}$ & -3.72030283 &
    d$_{2}$ & 1.072763 $\times$ 10$^{-1}$ \\
    a$_{3}$ & 0.2678 &
    b$_{3}$ & 5.148$\times$10$^{-5}$ &
    c$_{3}$ & 1.1412855 $\times$ 10$^{-1}$ &
    d$_{3}$ & 1.38385 $\times$ 10$^{-3}$ \\
    a$_{4}$ & -8.103$\times$10$^{-3}$ & & &
    c$_{4}$ & 2.093236 $\times$ 10$^{-5}$ & & \\
    a$_{5}$ & -2.079$\times$10$^{-5}$ & & & & & & \\
    \end{tabular}
    \caption{Empirically derived and model-fitted constants for calculating heat capacity and vaporization enthalpy of salt water, used in Eqs. \ref{eqn-L} and \ref{eqn-c_p}.}
    \label{table_cp-constants}
\end{center}
\end{table*}

\begin{figure}[t]
    \begin{center}
    \includegraphics[width=\textwidth]{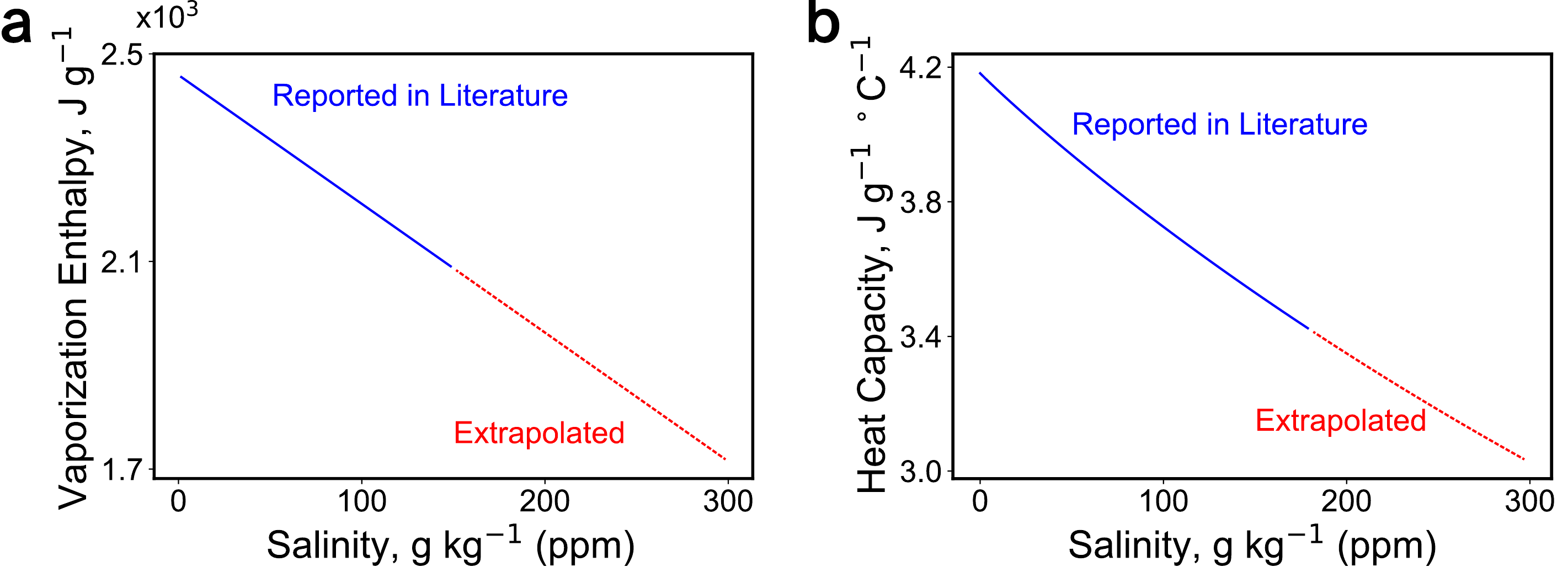}
    \end{center}
    \caption{(a) Heat capacity and (b) latent heat of brine as a function of salinity, plotted using Eqs. \ref{eqn-L}-\ref{eqn-c_p}, extrapolated from Ref. \cite{Generous2020} at NTP (20$^\circ$C and 0.1 MPa pressure)}
    \label{fig-heat_capacity}
\end{figure}

\noindent Here, $S$ is the salt concentration in g kg$^{-1}$ and T is the temperature in $^{\circ}$C. $a_{\rm n}$ are fitting constants of the Pitzer saltwater model, $b_{\rm n}$, $c_{\rm n}$, and $d_{\rm n}$ are empirically derived constants, and are listed in Table \ref{table_cp-constants}. We extrapolate these relations from the literature\cite{Generous2020} to higher salinity concentrations [Fig. \ref{fig-heat_capacity}]. Note that although there is an explicit pressure dependence in the literature, we have omitted it from Eq. \ref{eqn-c_p}, because it's a very weak effect ($\sim$10$^{0}$ J g$^{-1}$ $^\circ$C$^{-1}$ even when increasing to 100 atm pressure, and $\sim$10$^{-3}$ J g$^{-1}$ $^\circ$C$^{-1}$ for our experiments).

\par The majority of our measurements are of mass evaporation. Therefore, unless explicitly stated otherwise, any values of $Q$ throughout the following section are masses which have been converted using Eq. \ref{eqn-phase_change}. Note that Eqs. \ref{eqn-L} and \ref{eqn-c_p} yield a result in J kg$^{-1}$, and thus have been rescaled to J g$^{-1}$.

\par To isolate the evaporation that occurs from light absorption in the wick, we subtract the evaporation that occurs from increasing the specific surface area of the interface, i.e. from the enhancement that occurs by inserting a wick into water. From the normalized measurements, we estimate the efficiency of the light-enhanced evaporation.

We first use an integrating sphere to calculate the percent of light that is reflected to determine the light energy that is absorbed by the sample (Sec. S1). The absorbed light consists of two components: light that is absorbed by the AlN and light that is transmitted through the AlN and subsequently absorbed by the Al substrate. Both of these components contribute to heating of the reservoir if they do not enhance the evaporation rate. 

We calculate the energy-conversion efficiency, $\eta$, as

\begin{equation}
    \eta = \dfrac{Q_{\rm effective}}{Q_{\rm absorbed}}
    \label{eqn-eta}
\end{equation}

\noindent where $Q_{\rm effective}$ is the light energy used by the entire system [Eq. \ref{eqn-q_evap_wick}], and $Q_{\rm absorbed}$ is the total light energy absorbed by the sample over the course of an experiment [Eq. \ref{eqn-q_abs}]. $Q_{\rm effective}$, calculated below, accounts for the enhanced evaporation that occurs by increasing the specific surface area of the air-water interface.

\begin{equation}
    Q_{\rm effective} = Q_{\rm light}-Q_{\rm dark}
    \label{eqn-q_evap_wick}
\end{equation}

\noindent where $Q_{\rm light}$ and $Q_{\rm dark}$ are calculated from the mass-change measurements with the wick, with and without light. In the manuscript, we refer to $Q_{\rm dark}$ as the dark field control.

\begin{equation}
    Q_{\rm absorbed} = Pt(1-R_{\rm sample})
    \label{eqn-q_abs}
\end{equation}

\noindent $P$ is the measured power output of the LED, integrated over the steady-state time, $t$, of an experiment, at the location of the sample, and $R_{\rm sample}$ is the sample's measured reflectance [Eq. \ref{eqn-r_sample}]. Although the energy delivered to the sample is $\sim$6.2 kJ over an entire experiment, to account for the initial non-equilibrium conditions, we only consider energy and mass lost after $t=400$ minutes. In this case, $Pt$ is $\sim$3.1 kJ, and $Q_{\rm absorbed}$ is 2.1 - 2.3 kJ, depending on the wavelength of light. In Table \ref{table-efficiency_mass}, $m_{\rm 0}$ refers to the mass of water at $t=400$, rather than the absolute beginning of an experiment.

\begin{table*}[t]
    \begin{center}
    \rowcolors{2}{gray!25}{white}\begin{tabular}{ccccccc}
    \rowcolor{gray!10}
    $\lambda$,\, nm
    & $m_{\rm 0, \, light}$, g
    & $m_{\rm fin, \,light}$
    & $\Delta m_{light}$, g
    & $m_{\rm 0, \, dark}$, g
    & $m_{\rm fin, \,dark}$
    & $\Delta m_{dark}$, g\\
    \hline
    385
    & 35.1 - 41.5
    & 32.9 - 39.2
    & 1.9 - 2.3
    & 38.5 - 42.1
    & 37.4 - 41.4
    & 0.7 - 1.3 \\
    625
    & 35.9 - 42.0
    & 35.0 - 40.8
    & 0.8 - 1.2
    & 36.1 - 42.4
    & 35.5 - 41.3
    & 0.5 - 1.2 \\
    940
    & 36.7 - 51.5
    & 35.5 - 50.4
    & 0.9 - 1.2
    & 38.5 - 42.1
    & 37.4 - 41.4
    & 0.7 - 1.0 \\
    \end{tabular}
    \caption{Steady-state mass loss terms used to calculate energy efficiency [Eqs. \ref{eqn-phase_change}, \ref{eqn-eta}-\ref{eqn-q_abs}]. c = 3.0 J g$^{-1}$ $^{\circ}$C$^{-1}$ and L = 1.72$\times$ 10$^3$ J g $^{-1}$ are extrapolated from the literature\cite{Generous2020} [Eqs. \ref{eqn-L} and \ref{eqn-c_p}, Fig. \ref{fig-heat_capacity}].}
    \label{table-efficiency_mass}
    \end{center}
\end{table*}

As previously stated, the above calculations only approximate an efficiency, rather than provide an absolute measurement. A more accurate equation would be 

\begin{equation}
    \eta = \frac{Q_{\rm effective} - Q_{\rm waste,\, reservoir}}{Q_{\rm absorbed} - Q_{\rm waste,\, total}}
    \label{eqn-eta_waste}
\end{equation}

\begin{table*}[b]
    \begin{center}
    \rowcolors{2}{gray!25}{white}\begin{tabular}{cccccc}
    \rowcolor{gray!10}
    $\lambda$,\hspace{1mm}nm
    & $Q_{\rm light}$, kJ
    & $Q_{\rm dark}$, kJ
    & $Q_{\rm effective}$, kJ
    & $Q_{\rm absorbed}$, kJ
    & $\eta$, \% \\
    \hline
    385
    & 3.7 - 4.7
    & 1.4 - 2.5
    & 2.0 - 3.3
    & 2.2
    & 90 - 150\\
    625
    & 1.8 - 2.3
    & 1.3 - 2.0
    & 0.3 - 0.5
    & 2.3
    & 13-21 \\
    940
    & 1.5 - 2.4
    & 1.0 - 2.4
    & 0.0 - 0.6
    & 2.1
    & 1-27 \\
    \end{tabular}
    \caption{Energy utilization efficiency [Eq. \ref{eqn-eta}]. Calculated and synthesized from Table \ref{table-efficiency_mass} using Eq. \ref{eqn-phase_change}.}
    \label{table-efficiency_energy}
    \end{center}
\end{table*}

\noindent The actual efficiency range is expected to be slightly higher and slightly lower, depending on the ratios of $Q_{\rm effective}$/$Q_{\rm absorbed}$ and $Q_{\rm waste,\, reservoir}$/$Q_{\rm waste,\, total}$, as Eq. \ref{eqn-eta} does not account for ``waste heat,'' i.e. energy that is absorbed from the light, but which does not contribute to evaporation on the AlN surface.

$Q_{\rm waste,\, total} = Q_{\rm waste,\, reservoir} + \beta$, where $\beta = m_{\rm fin}c\Delta T$, and accounts for the temperature increase of the bulk water during an experiment. $\beta \ll Q_{\rm waste,\, reservoir}$, because of the relatively small $\Delta T$ term, which implies the following: $\frac{Q_{\rm effective}}{Q_{\rm absorbed}} < \frac{Q_{\rm effective-\beta}}{Q_{\rm absorbed-\beta}} \longrightarrow 
Q_{\rm effective}(Q_{\rm absorbed} - \beta) < Q_{\rm absorbed}(Q_{\rm effective} - \beta) \longrightarrow 
-(Q_{\rm effective})\beta < -(Q_{\rm absorbed})\beta \longrightarrow 
Q_{\rm effective} > Q_{\rm absorbed} \longrightarrow 
\eta > 1$

\noindent In other words, subtracting the same value from both the numerator and denominator of a fraction $>1$ will make the result larger, and it is easy to see that this relation is reversed in the opposite case. This means that our best-case efficiency for violet light is actually higher than 150\%, and our worst-case efficiency is lower than 90\%.

\par In Eq. \ref{eqn-eta}, we assume that all light is either reflected or extinguished by the AlN, and ignore the dynamics with the surrounding air, the Al substrate, and the bulk reservoir. Broadly speaking, the waste energy may leave the system in one of two ways: it can be absorbed by the Al substrate or internally reflected/waveguided by the AlN. In both cases, the waste energy may be radiated to the surrounding air if the gradient is steep enough, or into the water reservoir. Salt crystallization enthalpy is small compared to the vaporization phase-change energy.  Both return energy to the system during nucleation events. However, based on the thermodynamic and chemical properties of NaCl\cite{Chmarzynski1992}, the contribution of salt crystallization enthalpy is 30-50 J, compared to the $\sim$10$^{3}$ associated with light evaporation. The salt crystallization enthalpy is therefore ignored in calculations. The time-dependent interplay between these four interfaces creates a dynamic system which is difficult to characterize at present, especially in the case of porous media such as our AlN wick.

\section*{Sec. S4: Tables}

\begin{table*}[ht!]
\begin{center}
    \rowcolors{2}{white}{gray!20}
    \begin{tabular}{cccccc}
    \rowcolor{gray!10} & \multicolumn{5}{c}{Enhancement Rate, kg m$^{-2}$ h$^{-1}$} \\
    \hline
    \rowcolor{white}Salinity, wt\% & \multicolumn{5}{c}{Wavelength, nm (Intensity, mW cm$^{-2}$)}\\
    & Dark (0) & 385 (40) & 385 (130) & 625 (130) & 940 (130)\\
    \hline
    0 & 0.0 $\pm$ 0.0 & 0.09 $\pm$ 0.14 & 0.84 $\pm$ 0.24 & | & |\\
    9 & 0.0 $\pm$ 0.00 & 0.07 $\pm$ 0.12 & 0.34 $\pm$ 0.27 & | & |\\
    25 & 0.0 $\pm$ 0.17 & 0.30 $\pm$ 0.13 & 2.46 $\pm$ 0.79 & 0.15 $\pm$ 0.3 & 0.46 $\pm$ 0.36 \\
    \hline
    \end{tabular}
    \caption{Enhancement rate at different wavelengths and intensities of light.}
    \label{table-enhancement}
\end{center}
\end{table*}

\pagebreak

\newgeometry{margin=1cm}
\begin{sidewaystable*}[ht]
\begin{center}
    \rowcolors{2}{gray!20}{white}
    \begin{tabular}{cccccccc}
    \rowcolor{gray!10} Ref & Reported Efficiency & Above PT Limit? & Dark Field Sample? & Light Intensity & Mineralization Discussion & Temperature & Humidity\\
    \hline
    \cite{Haddad2021} & 50\% & No & No & 1 sun & Indirect & Yes & No\\
    \cite{Li2018} & None & No & No & 1 sun & Limited & Yes & Yes\\
    \cite{Li2018a} & None & No & No & 1 sun & Limited & Yes & Yes\\
    \cite{Bian2018} & 130\% & Yes & Yes, but unused & 1 sun & Yes & Yes & Limited\\
    \cite{Shi2021} & 119\% & Yes & Yes & 1 sun & No & Limited & Limited\\
    \cite{Wang2020} & >100\% & Yes & Yes & 1 sun & No & Yes & Limited\\
    \cite{Zhang2023} & 83.3\% & No & No & 1 and 1.5  sun & Limited & Yes & No\\
    \cite{Ma2019} & 96\% & No & No & 1 sun & Limited & Yes & No\\
    \cite{Liang2019} & 99.4\% & Yes & No & 1-4 sun & Yes & Yes & Limited\\
    \cite{Ito2015} & 80\% & No & No & 1 sun & No & Yes & Limited\\
    \cite{Guo2020} & 93\% & Yes & No & 1 sun & Yes & Yes & No\\
    \cite{Gao2021} & 179\% & Yes & Unreported & 1 sun & Limited & Yes & Limited\\
    \cite{Zhang2020b} & 97\% & Yes & Yes, unclear if used & 1 sun & Limited & Yes & No\\
    \cite{Dong2020} & 80\% & Yes & No & 1 sun & Limited & Yes & Limited\\
    \cite{Li2020} & Unreported & Yes & Yes & 1 sun & No & Yes & Limited\\
    \cite{Zhu2017} & 83\% & No & No & 10-sun & Limited & Yes & Unreported\\
    \cite{Zhu2018} & 67\% & No & No & 10 sun & Limited & Yes & No\\
    \cite{Qiao2019} & 92\% & No & No & 1-10 sun & No & Yes & No\\
    \cite{Kuang2019} & 75\% & No & Yes & 1-6 sun & Yes & No & No\\
    \cite{Zhang2020a} & 80\% & No & No & 1-3 sun & Yes & Yes & No\\
    \cite{Kumar2023} & Unreported & No & No & Unreported & No & Yes & No\\
    \cite{Javed2022} & 80\% & No & No & 1-2 sun & No & Yes & No\\
    \cite{Zhang2019} & 87\% & No & No & 1 sun & Limited & Yes & No\\
    \cite{Yang2018} & 87\% & No & Yes & 1-10 sun & No & Yes & Yes\\
    \cite{Shen2020} & 91\% & No & Yes & 1 sun & Yes & Yes & Limited\\
    \cite{Shi2018} & 79\% & No & Unreported & 1 sun & Yes & Yes & Limited\\
    Us & 85-113\% & No & Yes & 1 sun & Limited & Limited & Yes\\
    \end{tabular}
    \caption{Comparison to other work.}
    \label{comparison}
\end{center}
\end{sidewaystable*}
\restoregeometry

\section*{Sec. S5: Capillary Rise and Reusability Data}

\subsection*{Sec. S5.1: -OH Peak and Volumetric Flow Rate}

\begin{figure}[t]
    \begin{center}
    \includegraphics[width=\textwidth]{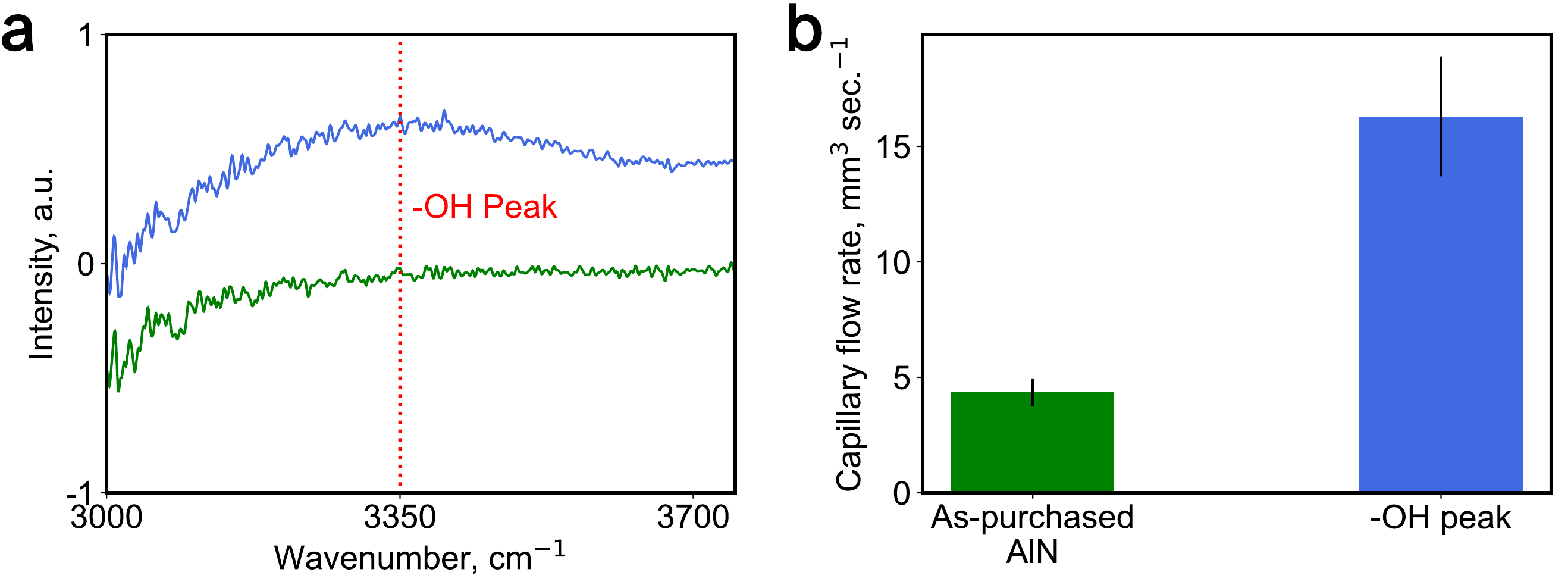}
    \end{center}
    \caption{(a) FTIR spectra and (b) capillary flow rate of AlN samples with and without an -OH peak.}
    \label{fig-al-oh-capillary}
\end{figure}

\begin{wrapfigure}[10]{r}{0.5\textwidth}
    \begin{center}
    \vspace{-15mm}
    \includegraphics[width=0.5\textwidth]{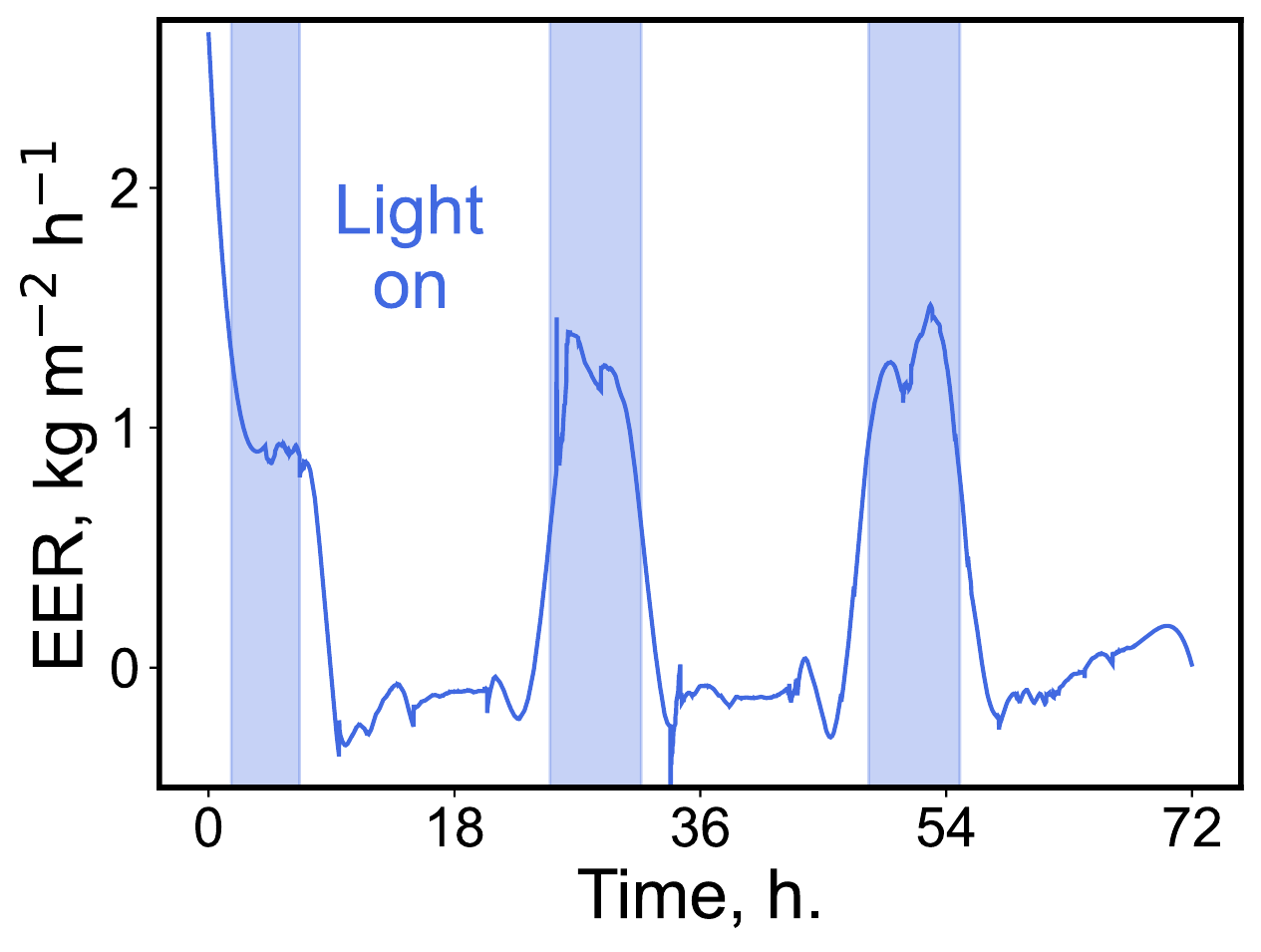}
    \end{center}
    \caption{72-hour experiment, in which violet light is cycled on and off to simulate overnight conditions.}
    \label{fig-72h-data}
\end{wrapfigure}

The surface -OH group enhances the capillary transport and water imbibition of water by our sample. First, we hydrolyze AlN powder in order to produce Al-OH groups within the network. We heat and agitate a slurry overnight with a magnetic stirring rod on a hot plate, set to 100 $^{\circ}$C. After heating and agitation, we confirm the presence of an -OH peak on the FTIR spectra. Once the -OH peak is confirmed, we perform water uptake rate measurements of both the hydrolyzed powder and the as-purchased AlN. We use a manual screw press to compact the powder into a 1 cm-diameter straw, then place the straw into a reservoir and measure the time until saturation. The sample which has an -OH peak in the FTIR spectra has a faster capillary rise time, which is $\sim$4$\times$ faster (16.3 mm$^{3}$ sec$^{-1}$ compared to 4.3 mm$^{3}$ sec$^{-1}$) than the as-purchased AlN [Fig. \ref{fig-al-oh-capillary}].

\subsection*{Sec. S5.2: Reusability and Sustained Enhancement}
To test the reusability and durability of our samples, we performed a 72-hour experiment where we cycled the light on and off to simulate day/night cycles. The results are shown in Fig. \ref{fig-72h-data}. The sample shows no visible sign of degradation after this time, and its evaporative performance remains within 10\% across all three days, with a median evaporation enhancement rate of 1.0 $\pm$ 0.1 kg m$^{-2}$ h$^{-1}$.

\bibliography{refs.bib}